\documentclass[runningheads]{llncs}

\usepackage[T1]{fontenc}
\usepackage{graphicx}
\usepackage{amsmath,amsfonts,amssymb}
\usepackage{mathrsfs}
\usepackage{algorithm}
\usepackage{algorithmic}
\usepackage{array}
\usepackage{subfig}
\usepackage{booktabs}
\usepackage{tabularx}
\usepackage{multirow}
\usepackage{xurl}
\usepackage{xcolor}
\usepackage{upgreek}
\usepackage{makecell}
\usepackage{listings}
\usepackage{marvosym}

\definecolor{codebg}{HTML}{F6F7F8}  
\definecolor{codekeyword}{HTML}{0000FF}
\definecolor{codecomment}{HTML}{008000}
\definecolor{codestring}{HTML}{A31515}

\def\BibTeX{{\rm B\kern-.05em{\sc i\kern-.025em b}\kern-.08em
    T\kern-.1667em\lower.7ex\hbox{E}\kern-.125emX}}
\begin{document}
\title{QUASAR: Quantum Satellite Architecture and Routing Simulator}

\author{
Yaliang Shi \and
Zi Wang \textsuperscript{(\Letter)} \and
Bangguo Yuan \and
Gaojie Wu \and
Zhiwei Zhao
}

\authorrunning{Y. Shi et al.}

\institute{
College of Computer Science and Engineering, University of Electronic Science and Technology of China, Chengdu 610051, China\\
\email{\{yaliang,wangzi,bangguo,gaojie,zhiwei\}@mobinets.org}
}
\maketitle
\setcounter{footnote}{0}
\begin{abstract}
The deployment of Low Earth Orbit (LEO) satellite constellations is an important step toward global-scale quantum networking. However, evaluating satellite quantum network protocols under spatiotemporal orbital dynamics and quantum physical constraints remains computationally expensive and challenging. In this paper, we propose QUASAR, a lightweight simulator for evaluating entanglement distribution in satellite-based quantum networks. QUASAR provides a decoupled architecture that integrates dynamic orbital topologies, time-varying optical transmittance, and quantum memory decoherence into network-layer attributes. To demonstrate its capabilities, we abstract and implement two representative hardware architectures: Simultaneous Downlink and On-Orbit Stitching. We further introduce an Entanglement Distribution Rate (EDR)-Aware Spatiotemporal Routing (EASR) heuristic as a reference workload. Our case study examines how QUASAR supports different satellite architectures, routing workloads, realistic orbital traces, concurrent requests, and scalable event-driven execution. With over \(85\%\) lower network-layer update latency than continuous polling, QUASAR provides a practical and extensible framework for future satellite quantum network protocol evaluation.

\keywords{Satellite Quantum Networks \and Entanglement Distribution \and Discrete-Event Simulation \and Spatiotemporal Routing \and Low Earth Orbit Constellations}
\end{abstract}
\section{Introduction}
The realization of a global quantum internet promises secure communication and distributed quantum computing~\cite{pan2024evolution}. To bridge isolated terrestrial quantum networks into a global infrastructure, satellite-terrestrial architectures have become increasingly important, with recent efforts outlining their use cases, architectures, and deployment roadmaps~\cite{de2023satellite}. Due to the exponential photon loss in terrestrial optical fibers, Low Earth Orbit (LEO) satellite constellations have emerged as a promising approach for global-scale entanglement distribution~\cite{yin2017satellite,goswami2025satellites}. These developments motivate systematic evaluation tools for satellite-based quantum network designs before large-scale deployment.

Current quantum network simulators target different design goals. Comprehensive discrete-event simulators such as NetSquid~\cite{coopmans2021netsquid} and SeQUeNCe~\cite{wu2021sequence} provide fine-grained models of quantum hardware, protocols, and control processes. They can capture photon-level timing and density-matrix state evolution, but such detail can incur substantial computational overhead for large-scale, dynamic LEO constellation evaluation. Network-layer simulators such as SimQN~\cite{chen2023simqn} improve scalability and ease of protocol evaluation. However, these engines are optimized for static, terrestrial topologies. They lack native support for continuous LEO orbital propagation, elevation-dependent atmospheric attenuation, and dynamic line-of-sight (LOS) contact graphs.

Consequently, current routing studies for satellite quantum networks often rely on customized optimization models and heuristics \cite{wei2024sky,chang2023entanglement,williams2024scalable,hu2024dynamic,gu2025quesat}. While these studies provide theoretical insights, they are usually evaluated under isolated or averaged physical conditions, making it difficult to capture the joint impact of LEO orbital dynamics, optical channel variation, and quantum memory decoherence. Evaluating routing paradigms ranging from instantaneous downlinks to memory-assisted spatial relays therefore requires an event-driven simulator that can abstract continuous physical constraints into dynamic network topologies and routing metrics.

To bridge the gap between static terrestrial network-layer simulators and the need for dynamic satellite quantum network evaluation, we propose QUASAR (Quantum Satellite Architecture and Routing Simulator).\footnote{Source code: \url{https://github.com/QuNets/quasar}.} 
QUASAR is implemented as a lightweight, decoupled Python overlay built on top of SimQN's discrete-event simulation core. It acts as a spatiotemporal abstraction bridge that processes LEO orbital geometry, optical transmittance, and memory decoherence. These physical states are exposed as dynamic network-layer attributes for event-driven routing evaluation.

The main contributions of this paper are threefold:
\begin{itemize}
\item We propose QUASAR, a lightweight spatiotemporal overlay built on top of SimQN. By decoupling continuous orbital mechanics and channel modeling from the discrete-event execution layer, QUASAR extends network-layer quantum simulation to dynamic LEO quantum network scenarios with reduced update overhead.
\item We abstract and formalize two representative quantum satellite architectures within the overlay---\emph{Simultaneous Downlink} and \emph{On-Orbit Stitching}---introducing temporal decoherence models to reflect the physical limits of quantum memories.
\item We conduct a system-level case study using Entanglement Distribution Rate (EDR)-Aware Spatiotemporal Routing (EASR) as a reference workload. The evaluation covers controlled architecture behavior, trace-driven orbital inputs, concurrent request workloads, and runtime mechanism ablation, demonstrating QUASAR's ability to support scalable satellite quantum network evaluation.
\end{itemize}

\section{Background and Related Work}
\label{sec:background_related}
This section reviews the quantum-networking primitives, simulators, satellite channel models, and routing studies that motivate QUASAR's reusable event-driven substrate for LEO satellite quantum network evaluation.

\subsection{Quantum Networking Primitives}
Unlike classical packet-switched networks, quantum networks do not primarily deliver data packets over a forwarding path. Instead, their fundamental network-layer service is to distribute high-fidelity Einstein--Podolsky--Rosen (EPR) pairs between remote users~\cite{shi2020concurrent,zhao2021redundant}. Such entangled pairs serve as consumable communication resources for higher-layer quantum protocols, including teleportation-based state transfer and distributed quantum applications.

\begin{figure}[t]
\centering
\includegraphics[width=0.8\linewidth]{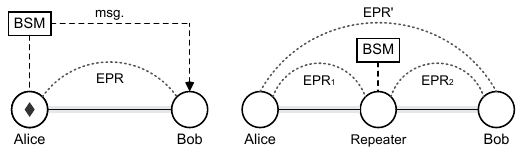}
\caption{Illustration of two quantum networking primitives: quantum teleportation on the left and entanglement swapping on the right.}
\label{fig:quantum_ops}
\end{figure}

Fig.~\ref{fig:quantum_ops} illustrates two basic primitives used in long-distance quantum networking. In quantum teleportation, an input qubit is transferred by consuming a shared EPR pair and a classical message produced after a Bell-state measurement (BSM). In entanglement swapping, two local EPR pairs are converted into a longer end-to-end EPR pair through an intermediate BSM at a repeater node. These primitives shift the network-layer objective from classical packet delivery to the establishment of usable end-to-end entanglement.

In satellite quantum networks, realizing these primitives is inherently spatiotemporal. Optical attenuation depends on satellite-ground distance, elevation angle, and link type, while feasible links appear only during intermittent contact windows. In memory-assisted operation, quantum memories temporarily store entangled qubits, which leads to storage-induced fidelity degradation. Therefore, evaluating satellite quantum-network protocols requires a simulator that can expose time-varying physical constraints as network-layer attributes. This observation motivates QUASAR's design: continuous orbital, channel, and memory dynamics are abstracted into dynamic topology and routing metrics for event-driven evaluation.
\subsection{Related Work}
\textbf{Quantum network simulators.}
Several quantum network simulators have been developed with different abstraction levels and design goals. NetSquid and SeQUeNCe provide comprehensive discrete-event frameworks for modeling quantum hardware, memories, protocol state machines, and control-plane behavior, making them suitable for detailed full-stack or hardware-aware evaluation~\cite{coopmans2021netsquid,wu2021sequence}. Other platforms such as SimulaQron, QuNetSim, and QuISP support quantum Internet software development, protocol experimentation, and architecture-level studies~\cite{dahlberg2018simulaqron,diadamo2021qunetsim,satoh2022quisp}. SimQN takes a complementary network-layer view by reducing quantum-state simulation overhead and supporting reusable network protocols~\cite{chen2023simqn}. These platforms have significantly advanced quantum-network experimentation, but they are not primarily designed around continuously evolving LEO satellite contact graphs, elevation-dependent optical attenuation, and event-driven routing recomputation under orbital dynamics.

\textbf{Satellite quantum channel modeling.}
A complementary line of work focuses on the physical feasibility and channel modeling of satellite quantum communication. Existing analyses characterize how propagation distance, contact geometry, atmospheric attenuation, and satellite-ground visibility affect achievable entanglement rates~\cite{khatri2021spooky,vasylyev2019satellite,klen2023numerical}. Recent work also considers Earth-to-satellite optical links and spatial diversity under realistic optical impairments~\cite{11015472,gonzalez2024satellite}. These studies provide important physical-layer models and feasibility insights, but they are not primarily designed as reusable network-layer simulation substrates for routing and event-driven evaluation.

\textbf{Satellite quantum network routing.}
Recent studies have begun to address routing, scheduling, and resource allocation problems in satellite-based quantum networks. Optimization-based approaches formulate satellite assignment, resource allocation, and path selection for entanglement distribution under satellite and ground-station resource constraints~\cite{wei2024sky}. Other works consider dynamic satellite movement, inter-satellite links, scalable scheduling policies, and space-ground integrated routing for satellite-assisted entanglement distribution~\cite{chang2023entanglement,williams2024scalable,hu2024dynamic}. More recent efforts further investigate satellite-assisted quantum Internet architectures for global-scale entanglement distribution~\cite{gu2025quesat}. These studies provide important algorithmic insights, but reusable event-driven substrates for comparing satellite quantum-network architectures, dynamic channel attributes, and routing workloads under a common abstraction remain less explored.

\textbf{Position of QUASAR.}
QUASAR is positioned between detailed physical layer simulators and customized satellite routing optimization studies. Rather than replacing full stack quantum network simulators or high fidelity optical channel models, QUASAR provides a lightweight spatiotemporal abstraction layer for satellite entanglement distribution evaluation. It integrates orbital visibility, optical transmittance, and memory related fidelity degradation into dynamic network layer attributes, allowing different satellite quantum network architectures and routing workloads to be evaluated under a common event driven execution model. This design makes QUASAR complementary to existing tools by preserving network-layer programmability while retaining key time-varying LEO constraints.

\section{The QUASAR Simulator Design}
\label{sec:design}
\begin{figure}[t] 
    \centering
    \includegraphics[width=\linewidth]{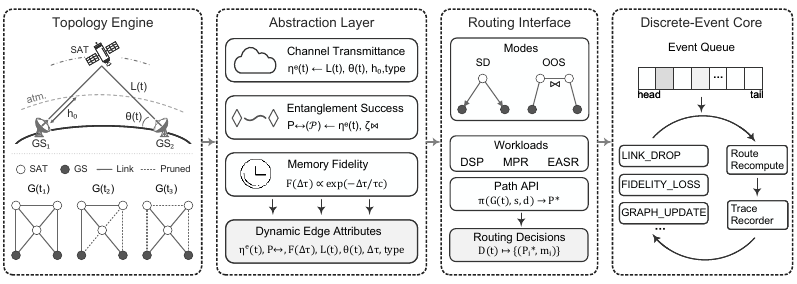}
    \caption{Core simulation modules of the QUASAR framework.}
    \label{fig:architecture}
\end{figure}

QUASAR is a decoupled Python overlay that bridges continuous physical calculations and discrete-event network simulation. As illustrated in Fig.~\ref{fig:architecture}, it abstracts orbital visibility, channel transmittance, and memory decoherence into evolving network-layer attributes consumed by the simulation engine.

\subsection{Spatiotemporal Topology Engine}
\label{subsec:topology}
The foundation of QUASAR is its spatiotemporal topology engine, designed to natively model the high-velocity mobility of LEO constellations. The engine supports configurable Walker-Delta constellations and standard Two-Line Element (TLE) sets, and utilizes the SGP4 orbital propagator to track satellite trajectories in the Earth-Centered Inertial (ECI) coordinate frame~\cite{11015472}. At each $\Delta t = 100$ ms simulation slot, the physical engine evaluates orbital visibility, channel transmittance, and memory states, and then triggers network-layer updates only when the corresponding visibility, channel, or fidelity conditions change.

At any given time $t$, for a satellite located at position vector $\mathbf{r}_s(t)$ and a ground station at $\mathbf{r}_g$, the engine calculates the instantaneous Euclidean slant range $L(t) = \| \mathbf{r}_s(t) - \mathbf{r}_g \|_2$ and the respective elevation angle $\theta(t)$. To prevent the underlying routing algorithms from attempting infeasible connections, the engine enforces a Line-of-Sight (LOS) visibility mask. It proactively prunes the network graph by dropping any space-to-ground edge where $\theta(t)$ falls below a minimum hardware-dictated threshold $\theta_0$, capturing Earth blockage and low-elevation visibility constraints. For inter-satellite links, QUASAR similarly updates candidate edges according to mutual visibility and hardware range constraints. An ISL is included in \(\mathcal{E}_t\) only when the corresponding satellite pair has an unobstructed optical path and satisfies the configured range limit; otherwise, the edge is pruned before being exposed to the routing layer.

\subsection{Dynamic Channel Models}
\label{subsec:channel}
Quantum signals traversing satellite networks experience distance-dependent attenuation, which directly bounds the entanglement generation rate. To capture this spatial heterogeneity, the simulator differentiates the transmittance of a given edge $e$ based on its physical environment: satellite-to-ground links (SGLs) penetrating the atmosphere, and inter-satellite links (ISLs) operating in the vacuum of space. Before applying environmental penalties, QUASAR computes a baseline free-space transmittance for each edge. This baseline captures propagation loss induced by the instantaneous propagation distance, telescope apertures, and beam divergence. The total edge transmittance, denoted as $\eta^{(e)}(t)$, is then instantiated as
\begin{equation}
\eta^{(e)}(t)=
\begin{cases}
\eta_0^{(e)}(t)\cdot \kappa,
& e \in \mathcal{E}_{\mathrm{ISL}}(t), \\[1mm]
\eta_0^{(e)}(t)\cdot \eta_\alpha(\theta_e(t))\cdot \kappa,
& e \in \mathcal{E}_{\mathrm{SGL}}(t),
\end{cases}
\label{eq:total_transmittance}
\end{equation}
where \(\eta_0^{(e)}(t)\) denotes the edge-specific free-space transmittance and \(\kappa\) absorbs fixed optical implementation losses such as coupling loss, detector efficiency, internal optical transmittance, and residual pointing loss. For SGLs, QUASAR further applies an elevation-dependent atmospheric penalty. As the satellite approaches the horizon, the optical path length through the atmosphere increases. We use the standard cosecant-law approximation \cite{klen2023numerical}:
\begin{equation}
\eta_\alpha(\theta_e(t)) =
\exp\left( -\frac{\alpha h_0}{\sin \theta_e(t)} \right),
\label{eq:cosecant_law}
\end{equation}
where $\alpha$ is the linear atmospheric attenuation coefficient, $h_0$ denotes the effective thickness of the atmosphere, and $\theta_e(t)$ is the instantaneous elevation angle of edge $e$. The resulting $\eta^{(e)}(t)$ is exposed as a dynamic edge attribute for network-layer pathfinding.

\subsection{Loss and Fidelity Model}
\label{subsec:loss}
Beyond spatial photon loss (erasure errors), quantum networks suffer from operational penalties and temporal decoherence (depolarizing errors). To decouple these complex quantum physical constraints into classical network-layer heuristics, QUASAR employs a lightweight success-probability and fidelity tracking model.

In the \emph{On-Orbit Stitching} paradigm, multi-hop entanglement distribution requires successive Bell state measurements (BSMs). We denote the probabilistic success rate of the $k$-th orbital relay as $\zeta_{\bowtie}^{(k)}$. The end-to-end success probability over a spatial path \(\mathcal{P}\) is modeled as
\begin{equation}
    \mathbb{P}_{\leftrightarrow}(\mathcal{P}) = \prod_{e \in \mathcal{P}} \eta^{(e)} \prod_{k \in \mathcal{V}_{\bowtie}} \zeta_{\bowtie}^{(k)},
    \label{eq:e2e_prob}
\end{equation}
where $\mathcal{V}_{\bowtie}$ is the subset of intermediate routing satellites performing entanglement swapping.

Furthermore, to model temporal decoherence during the orbital carry phase, let $\tau_c$ be the characteristic coherence time of the onboard quantum memory. Given an initial heralded fidelity $\mathcal{F}_0$, the degraded fidelity \(\mathcal{F}(\Delta \tau)\) after a physical storage duration \(\Delta \tau\) is modeled as
\begin{equation}
    \mathcal{F}(\Delta \tau) = \frac{1}{4} + \left( \mathcal{F}_0 - \frac{1}{4} \right) \exp\left( -\frac{\Delta \tau}{\tau_c} \right).
    \label{eq:decoherence}
\end{equation}

Instead of hardcoding a specific network policy, QUASAR decouples the physical simulation from the algorithmic execution layer. The simulator continuously tracks the geometric survival probability $\mathbb{P}_{\leftrightarrow}$ and the temporal fidelity $\mathcal{F}(\Delta \tau)$, exposing them as dynamic physical attributes for routing evaluation. Any higher-layer protocol evaluated within the framework can enforce the fidelity threshold condition $\mathcal{F}(\Delta \tau) \ge \mathcal{F}^\ast$. By exposing these physical attributes, QUASAR allows the network layer to balance the spatial probabilistic penalty of multi-hop swapping against the temporal decoherence induced by memory buffering.

\section{Satellite Architectures and Routing}
\label{sec:arch_routing}
Operating on top of the physical models abstracted in Section \ref{sec:design}, the QUASAR overlay maintains a dynamic topology $\mathcal{G}(t) = (\mathcal{V}, \mathcal{E}_t)$ and exposes dynamic edge attributes for network-layer evaluation. To capture representative LEO quantum-network capabilities, QUASAR supports two operational paradigms.

\subsection{Quantum Hardware Architectures}
\label{subsec:hardware}
The routing decision space is fundamentally constrained by the satellite's onboard hardware, specifically the presence and quality of quantum memories.

\paragraph{Simultaneous Downlink (SD).}
This paradigm assumes satellites act purely as entangled photon-pair sources (e.g., via SPDC) without memory buffering \cite{goswami2023satellite}. A distribution opportunity is valid only when a single satellite simultaneously maintains Line-of-Sight (LOS) to both target nodes. Consequently, the temporal delay is zero ($\Delta \tau = 0$), bypassing decoherence but restricting the spatial routing space to highly constrained geometrical intersections.

\paragraph{On-Orbit Stitching (OOS).}
Satellites equipped with quantum memories can buffer incoming qubits and act as memory-assisted relays. This decoupling allows entanglement generation and subsequent swapping ($\bowtie$) to occur asynchronously within feasible contact opportunities. However, memory-assisted operation incurs an inevitable fidelity penalty. Recalling \eqref{eq:decoherence}, the penalty is governed by the coherence time $\tau_c$. For a path $\mathcal{P}$ accumulating a total storage duration $\Delta \tau(\mathcal{P}) = \sum_{v \in \mathcal{V}_{\bowtie}} \Delta \tau_v$, the routing protocol must account for the exponential decoherence of fidelity.

These hardware paradigms expose different decision spaces in QUASAR. Under SD, routing reduces to selecting a satellite that simultaneously covers both target nodes; if no such satellite exists, the request is infeasible. Under OOS, routing becomes a memory-assisted spatiotemporal path-selection problem over $\mathcal{G}(t)$.

\subsection{Routing Interface and Reference Workloads}
\label{subsec:workloads}
Operating at the boundary of this physical-to-network decoupling, the QUASAR routing interface ingests the physical attributes generated by the models in Section~\ref{sec:design}. It exposes a dynamic topology whose edges carry real-time attributes such as transmittance \(\eta^{(e)}(t)\), edge type, and visibility status, while OOS-based routing additionally queries predicted LOS contact windows to determine feasible stitching opportunities and accumulated memory-residence delay \(\Delta \tau\).

To provide simple reference points, QUASAR includes two routing baselines representing opposite extremes in the decision space. Dynamic Shortest Path (DSP) minimizes instantaneous physical hops without explicitly accounting for optical loss, while Max-Probability Routing (MPR) greedily selects high-transmittance paths without explicitly accounting for temporal memory decoherence.

As a concrete reference workload, QUASAR includes the EDR-Aware Spatiotemporal Routing (EASR) heuristic, which consumes the dynamic attributes exposed by the event-driven interface to evaluate the joint effect of spatial loss and temporal decoherence. For memory-assisted path selection, we formulate the reference EDR-aware objective as
\begin{equation}
    \mathcal{P}^\ast = \arg\max_{\mathcal{P}} \left[ \mathbb{P}_{\leftrightarrow}(\mathcal{P}) \cdot \exp\left( -\frac{\Delta \tau(\mathcal{P})}{\tau_c} \right) \right].
    \label{eq:objective}
\end{equation}
To integrate this multiplicative objective into an efficient discrete-event pathfinding structure, we apply a negative logarithmic transformation, yielding a unified time-dependent edge weight
\begin{equation}
    \omega_{u,v}(t) = -\ln \eta^{(u,v)}(t) - \ln \zeta_{\bowtie}^{(v)} + \frac{\Delta \tau_v}{\tau_c}.
    \label{eq:edge_weight}
\end{equation}
Here, $\Delta \tau_v$ denotes the memory storage time at repeater node $v$ in a feasible OOS stitching opportunity, capturing the local buffering and coordination delay before entanglement swapping.

This formulation can be specialized to the selected hardware abstraction. Under the memoryless SD paradigm, it reduces to selecting a feasible simultaneous dual-downlink opportunity based on the joint instantaneous downlink transmittance, since $\Delta\tau=0$ and no memory-assisted swapping is performed. Under OOS, it balances multi-hop spatial penalties against temporal decoherence penalties.

\renewcommand{\algorithmicrequire}{\textbf{Input:}}
\renewcommand{\algorithmicensure}{\textbf{Output:}}
\begin{algorithm}[t]
\caption{EDR-Aware Spatiotemporal Routing (EASR)}
\label{alg:easr}
\begin{algorithmic}[1]
\REQUIRE 
\begin{tabular}[t]{@{}l@{}}
Dynamic graph $\mathcal{G}(t)$, Source $s$, Destination $d$, Coherence time $\tau_c$, \\Fidelity threshold $\mathcal{F}^\ast$
\end{tabular}
\ENSURE EDR-aware feasible path $\mathcal{P}^\ast$
\STATE Initialize cost array $\mathcal{C}[v] \leftarrow \infty$ for all $v \in \mathcal{V}$
\STATE Initialize Priority Queue $\mathcal{Q}$, insert $(s, 0)$
\STATE $\mathcal{C}[s] \leftarrow 0$
\WHILE{$\mathcal{Q}$ is not empty}
    \STATE Extract node $u$ with minimum $\mathcal{C}[u]$ from $\mathcal{Q}$
    \IF{$u == d$}
        \RETURN Reconstruct Path $\mathcal{P}^\ast$
    \ENDIF
    \FOR{each neighbor $v$ of $u$ in $\mathcal{G}(t)$}
        \STATE $\widehat{\tau} \leftarrow \Delta\tau(s\leadsto u\to v)$ \COMMENT{Cumulative delay}
        \STATE Estimate $\mathcal{F}(\widehat{\tau})$ using \eqref{eq:decoherence}
        \IF{$\mathcal{F}(\widehat{\tau}) < \mathcal{F}^\ast$}
            \STATE \textbf{continue} \COMMENT{Prune paths violating fidelity}
        \ENDIF
        \STATE Compute edge weight $\omega_{u,v}(t)$ using \eqref{eq:edge_weight}
        \IF{$\mathcal{C}[u] + \omega_{u,v}(t) < \mathcal{C}[v]$}
            \STATE $\mathcal{C}[v] \leftarrow \mathcal{C}[u] + \omega_{u,v}(t)$
            \STATE Insert $(v, \mathcal{C}[v])$ into $\mathcal{Q}$
        \ENDIF
    \ENDFOR
\ENDWHILE
\RETURN $\emptyset$ \COMMENT{No feasible path found}
\end{algorithmic}
\end{algorithm}
Algorithm~\ref{alg:easr} illustrates how the EASR workload consumes QUASAR's dynamic edge attributes. The priority queue expands the partial path with the lowest accumulated spatiotemporal cost. For each candidate extension, the algorithm estimates the cumulative storage delay $\widehat{\tau}$ and prunes the path if the resulting fidelity violates $\mathcal{F}^\ast$. Feasible extensions are then ranked by the edge weight in \eqref{eq:edge_weight}, which combines optical loss, swapping success, and temporal storage cost. In this design, fidelity acts as a hard feasibility filter, while the edge weight provides a soft ranking among feasible paths.

\section{Software Architecture and Implementation}
\label{sec:software_arch}
Building on the physical models in Section~\ref{sec:design} and the routing abstractions in Section~\ref{sec:arch_routing}, this section presents three software interfaces of QUASAR: topology instantiation, event-driven physical decoupling, and custom routing orchestration. The listings illustrate how users configure and run QUASAR workloads through these interfaces.

\subsection{Spatiotemporal Topology Instantiation}
\label{subsec:sw_topology}
The foundation of any QUASAR simulation is the instantiation of the physical network overlay. Rather than manually defining static adjacency matrices, researchers can construct large-scale Low Earth Orbit (LEO) constellations natively using standard orbital specifications.

As demonstrated in Listing \ref{code:topology}, QUASAR abstracts physical entities into object-oriented \texttt{Node} classes. Ground stations are instantiated via exact geographical coordinates, while the dynamic satellite nodes are generated from orbital specification files, such as TLE datasets or configurable Walker-Delta constellation descriptions. The framework then establishes dynamic quantum connections with time-varying visibility and channel attributes.

\begin{lstlisting}[caption={Instantiation of LEO constellation and ground stations.}, label={code:topology}]
# 1. Topology Generation: Satellites and Stations
import quasar

# Initialize core and add Ground Station (GS)
net_core = quasar.Network.get_instance()
gs_pek = quasar.GroundStation(
    "GS_PEK", lat=39.9, lon=116.4)
net_core.add_node(gs_pek)

# Load LEO constellation config
leo_nodes = quasar.Constellation("walker.txt")
net_core.add_nodes(leo_nodes)

# Establish dynamic quantum channels
for sat in leo_nodes:
    net_core.add_q_connection(
        sat, gs_pek, dynamic_loss=True)
\end{lstlisting}

\subsection{Event-Driven Physical Decoupling}
\label{subsec:sw_event_bridge}
A primary challenge in evaluating LEO quantum networks is the computationally prohibitive physical layer. Existing simulators either rely on static physical snapshots or execute continuous environmental polling, both of which impose severe limits on scalability.

QUASAR introduces a \texttt{Discrete-Event Bridge}. As shown in Listing~\ref{code:bridge}, the physical engine triggers state updates only when visibility, range, or fidelity thresholds are crossed, keeping spatial calculations decoupled from the network-layer event loop.

\begin{lstlisting}[caption={Event-driven bridge decoupling physics from discrete events.}, label={code:bridge}]
# 2. Core Implementation: Event-Driven Bridge
class EventBridge:
    def check_thresholds(self, edge_st):
        # Monitor physical attributes
        if not edge_st.available:
            self.trigger(
                "LINK_DROP", edge_st.nodes)
        elif edge_st.fidelity < F_STAR:
            self.trigger(
                "FIDELITY_LOSS", edge_st.nodes)
        else:
            # Update additive spatiotemporal weight
            w = -math.log(edge_st.eta) + \
                (edge_st.dt / TAU_C)
            self.update_weight(edge_st.nodes, w)
\end{lstlisting}

\subsection{Custom Orchestration Interface}
\label{subsec:sw_orchestration}
The primary objective of QUASAR is to serve as a versatile testbed for advanced network-layer algorithms. By exposing a unified \texttt{RoutingAlgo} interface, the framework allows researchers to evaluate complex spatiotemporal heuristics without manually managing the underlying orbital and quantum optical dynamics.

To illustrate this extensibility, Listing \ref{code:routing} demonstrates the integration of the EDR-Aware Spatiotemporal Routing (EASR) reference workload. The orchestrator dynamically calculates the unified edge weight---combining the negative logarithm of spatial success probability with the temporal decoherence penalty---and executes a modified shortest-path search. 

\begin{lstlisting}[caption={Custom orchestrator interface demonstrating EASR integration.}, label={code:routing}]
# 3. Custom Orchestrator: EDR-Aware Routing
class EASR_Orchestrator(quasar.RoutingAlgo):
    def compute_route(self, src, dst, graph):
        # EASR: Minimize spatiotemporal cost
        def easr_cost(edge):
            p_succ = edge.eta * edge.swap_prob
            dec_penalty = edge.dt / TAU_C
            return -math.log(p_succ) + dec_penalty

        # Execute dynamic shortest path search
        path = graph.dijkstra(
            src, dst, weight_fn=easr_cost)
        return path

# Usage: Register heuristic and execute
overlay.set_algorithm(EASR_Orchestrator())
overlay.run(duration=86400, dt=0.1)
\end{lstlisting}

\section{Case Study and System Evaluation}
\label{sec:evaluation}

We evaluate QUASAR through a case study covering controlled architecture behavior, routing workload sensitivity, event-driven runtime overhead, trace-driven orbital inputs, concurrent OD workloads, and runtime mechanism ablation. Together, these experiments examine QUASAR's ability to expose physical dynamics as network-layer attributes and to support scalable satellite quantum-network evaluation.

\subsection{Simulation Setup}
\label{subsec:setup}

The simulation is driven by the QUASAR Python overlay integrated with a discrete-event network core. We use a configurable Walker-Delta LEO constellation to provide repeated and reproducible global contact opportunities. Unless otherwise specified, the controlled studies use a fixed-size Walker-Delta constellation with $N=60$ satellites to isolate architecture- and memory-related effects. The runtime scalability benchmark varies the constellation size from $20$ to $800$ satellites, and the runtime ablation is conducted at the largest constellation setting. QUASAR also supports trace-driven orbital inputs through TLE files; in the trace-driven topology experiment, we use a $60$-satellite Starlink TLE subset obtained from CelesTrak as a realistic LEO orbital trace, not as a quantum communication deployment~\cite{celestrak_gp}. Satellite positions are propagated with a time resolution of $\Delta t = 100$ ms, and the resulting visibility, channel transmittance, and memory states are exposed to the network layer as dynamic attributes. The baseline physical parameters, inspired by near-term quantum satellite studies and analytical bounds~\cite{meister2025simulation,yehia2024connecting}, are summarized in Table~\ref{tab:sim_params}.

Our evaluation uses three workload settings. First, for the controlled architecture and workload-sensitivity studies, we isolate a representative long-distance ground-station pair between Houston ($\mathrm{GS}_1$) and Washington, DC ($\mathrm{GS}_2$)~\cite{khatri2021spooky}. This single-pair setting allows us to separate the effects of orbital visibility, optical loss, and memory decoherence from traffic-matrix effects. Second, for runtime scalability and mechanism ablation, we use a macro-benchmark with up to $32$ worldwide ground stations and continuously sampled origin-destination requests, so that the simulator is stressed by increasing constellation size and routing activity. Third, for the concurrent OD workload study, we fix the Walker-Delta constellation at $N=60$ satellites, use the same $32$-station macro-benchmark setting, and increase the number of concurrent OD pairs from $4$ to $32$. We report aggregate EDR across all active requests in ebits/s, where EDR is computed from the source repetition rate \(R_0\), the end-to-end path success probability, and fidelity-threshold filtering.

\begin{table}[t]
\caption{Baseline Simulation Parameters}
\label{tab:sim_params}
\centering
\begin{tabular}{@{}llc@{}}
\toprule
\textbf{Parameter} & \textbf{Symbol} & \textbf{Value} \\
\midrule
\multicolumn{3}{@{}l}{\textit{Orbital Dynamics}} \\
Constellation Type & -- & Walker-Delta \\
Orbit Altitude & -- & 500~km \\
Inclination Angle & -- & $53^\circ$ \\
\midrule
\multicolumn{3}{@{}l}{\textit{Quantum Channels}} \\
Base Transmittance (at 500 km) & $\eta_0$ & $10^{-3}$ \\
Source Repetition Rate & $R_0$ & $10^8$~s$^{-1}$ \\
Atmospheric Attenuation & $\alpha$ & 0.01~km$^{-1}$ \\
Effective Atm. Thickness & $h_0$ & 20~km \\
Static Hardware Efficiency & $\kappa$ & 0.85 \\
\midrule
\multicolumn{3}{@{}l}{\textit{Hardware \& Routing Constraints}} \\
Swap Success Probability & $\zeta_{\bowtie}$ & 0.60 \\
Initial Heralded Fidelity & $\mathcal{F}_0$ & 0.99 \\
Target Fidelity Threshold & $\mathcal{F}^\ast$ & 0.75 \\
Baseline Coherence Time & $\tau_c$ & 100~ms \\
Minimum LOS Elevation & $\theta_0$ & $15^\circ$ \\
\bottomrule
\end{tabular}
\end{table}

\subsection{Controlled Mechanism Studies}
\label{subsec:controlled_mechanisms}

\begin{figure}[!t]
\centering
\subfloat[Architecture.]{
\includegraphics[width=0.45\linewidth]{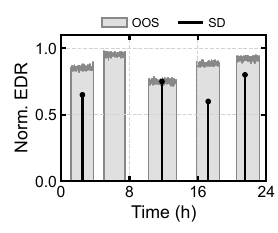}
\label{fig:res_arch}
}
\hfill
\subfloat[Workload.]{
\includegraphics[width=0.45\linewidth]{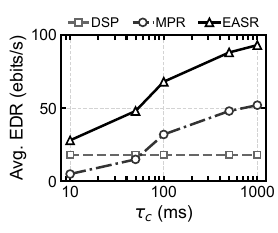}
\label{fig:res_workload}
}
\caption{Controlled mechanism studies enabled by QUASAR. (a) Architecture-dependent EDR under SD and OOS. (b) Routing workload sensitivity to memory coherence time.}
\label{fig:controlled_results}
\end{figure}

We first examine how QUASAR represents the service opportunities exposed by different hardware abstractions. Fig.~\ref{fig:controlled_results}(a) shows the normalized instantaneous EDR between $\mathrm{GS}_1$ and $\mathrm{GS}_2$ over a $24$-hour simulation horizon.

Under the Simultaneous Downlink (SD) paradigm, feasible service appears only at sparse instants when a single satellite simultaneously covers both ground stations. These short pulse-like opportunities reflect the geometric constraint of memoryless dual downlinking. Under the On-Orbit Stitching (OOS) abstraction, onboard quantum memories allow entanglement generated across separate contact intervals to be buffered and stitched later, producing wider feasible EDR windows. This result illustrates how QUASAR separates hardware-dependent feasibility from the routing workload: the same physical scenario exposes different network-layer opportunities depending on the selected architecture model.

We next use QUASAR's workload interface to compare routing behavior under different memory-coherence regimes. We sweep the quantum memory coherence time $\tau_c$ from $10$ ms to $1$ s and instantiate three reference workloads: Dynamic Shortest Path (DSP), Max-Probability Routing (MPR), and EDR-Aware Spatiotemporal Routing (EASR). As shown in Fig.~\ref{fig:controlled_results}(b), the three workloads respond differently to the same physical constraints. DSP primarily minimizes hop count and therefore remains insensitive to $\tau_c$, but its average EDR remains lower because it does not explicitly account for optical loss or memory decoherence. MPR improves link quality by favoring higher-transmittance paths, but its performance is limited when memories decohere rapidly. EASR achieves the highest EDR among the reference workloads because it combines spatial loss and temporal storage penalties through the edge weight in \eqref{eq:edge_weight}. Rather than serving as a standalone algorithmic benchmark, this comparison demonstrates that QUASAR can host multiple routing workloads and expose their sensitivity to the same underlying physical model.

\subsection{Event-Driven Runtime Behavior}
\label{subsec:event_runtime}

\begin{figure}[!t]
\centering
\subfloat[Events.]{
\includegraphics[width=0.45\linewidth]{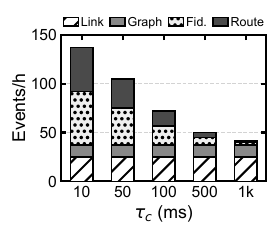}
\label{fig:res_event}
}
\hfill
\subfloat[Runtime.]{
\includegraphics[width=0.45\linewidth]{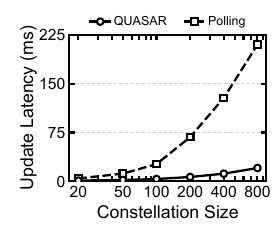}
\label{fig:res_runtime}
}
\caption{Event-driven runtime behavior of QUASAR. (a) Triggered-event composition under different memory coherence times. (b) Network-layer update latency under increasing constellation size.}
\label{fig:event_runtime_results}
\end{figure}

Fig.~\ref{fig:event_runtime_results}(a) reports the triggered-event composition recorded by the discrete-event core under the same controlled workload setting. Topology-driven events, including link and graph updates, remain stable across different values of $\tau_c$ because they are mainly determined by orbital visibility. In contrast, fidelity-loss events decrease as $\tau_c$ increases, since longer memory lifetimes reduce decoherence-induced violations. The number of route recomputation events follows the same trend, indicating that fewer physical-state violations propagate to the routing layer when memory coherence improves. This event-level view illustrates how QUASAR converts continuous orbital and decoherence dynamics into discrete updates consumed by network-layer workloads.

We then evaluate QUASAR's runtime overhead as the constellation size increases from $20$ to $800$ satellites. To stress the simulation engine beyond the single-pair case study, we use a macro-benchmark with $32$ worldwide ground stations and a continuous stream of randomly sampled origin-destination requests. At each $\Delta t = 100$ ms simulation slot, the physical engine evaluates orbital visibility, channel transmittance, and memory states, while network-layer attributes and routing states are refreshed only when the corresponding visibility, channel, or fidelity events are triggered.

Fig.~\ref{fig:event_runtime_results}(b) compares QUASAR's event-driven update mechanism with a continuous polling baseline. Both settings perform the same slot-level physical evaluation of orbital visibility, channel transmittance, and memory states; the reported latency measures downstream network-layer updates, including edge-attribute refreshes and routing recomputation. The polling baseline refreshes candidate link and routing states at every slot, whereas QUASAR triggers such updates only when visibility, channel, or fidelity changes produce relevant network events. Together with visibility and range pruning, this event filtering keeps the exposed graph sparse and reduces network-layer update latency by more than $85\%$ at the largest constellation setting.

\subsection{System Macro-Benchmarks}
\label{subsec:macro_benchmarks}

To evaluate QUASAR beyond the controlled single-pair setting, we extend the case study to trace-driven orbital inputs and concurrent OD routing workloads.
\begin{figure}[!t]
\centering
\subfloat[Trace.]{
\includegraphics[width=0.45\linewidth]{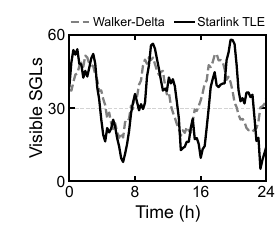}
\label{fig:trace}
}
\hfill
\subfloat[Concurrent OD.]{
\includegraphics[width=0.45\linewidth]{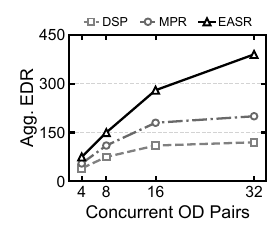}
\label{fig:multiod}
}
\caption{Trace-driven and concurrent-workload evaluation. (a) Visibility dynamics under Walker-Delta and Starlink TLE inputs. (b) Aggregate EDR under increasing concurrent OD pairs.}
\label{fig:stress_tests}
\end{figure}

We first demonstrate QUASAR's spatiotemporal topology engine using a Starlink TLE trace as a realistic LEO orbital input. Fig.~\ref{fig:stress_tests}(a) compares the instantaneous number of visible satellite-to-ground links over a $24$-hour horizon. Compared with the smoother periodic visibility pattern generated by the controlled Walker-Delta configuration, the Starlink TLE trace exhibits more irregular contact dynamics. This result shows that QUASAR can ingest non-ideal orbital traces and drive the same visibility, channel, and event pipeline used in the controlled experiments.

Fig.~\ref{fig:stress_tests}(b) further stresses the routing workload interface by increasing the number of concurrent origin-destination (OD) pairs from $4$ to $32$. The aggregate EDR is reported in ebits/s across all active OD pairs. As concurrency increases, all workloads deliver higher aggregate EDR, but their scaling behaviors differ because they respond differently to spatial channel loss, contact availability, and memory decoherence. DSP and MPR show earlier saturation under heavier request concurrency, while EASR sustains a higher aggregate EDR by accounting for both spatial and temporal costs. The aggregate EDR in Fig.~\ref{fig:stress_tests}(b) is measured across randomly sampled concurrent OD pairs and is therefore not directly comparable to the single-pair EDR in Fig.~\ref{fig:controlled_results}(b). This experiment is not intended as a standalone algorithmic benchmark or a full resource-contention scheduling model; rather, it demonstrates that QUASAR can drive multiple routing workloads under concurrent OD traffic and expose how aggregate service rates evolve with request concurrency.

\subsection{Runtime Mechanism Ablation}
\label{subsec:runtime_ablation}

Finally, Table~\ref{tab:runtime_ablation} reports a runtime mechanism ablation on the $800$-satellite benchmark. We start from a continuous polling baseline and then progressively activate three mechanisms: visibility pruning, channel event filtering, and fidelity event filtering. Visibility pruning reduces the number of infeasible satellite-ground and inter-satellite candidates exposed to the routing layer. Channel event filtering avoids refreshing network-layer edge attributes unless channel changes produce relevant events. Fidelity event filtering further reduces redundant route recomputation caused by memory-state updates.

\begin{table}[t]
\centering
\footnotesize
\setlength{\tabcolsep}{3.8pt}
\caption{Runtime Mechanism Ablation}
\label{tab:runtime_ablation}
\begin{tabular}{@{} l c c c r r r @{}}
\toprule
\multirow{2}{*}{\textbf{Config.}}
& \multicolumn{3}{c}{\textbf{Mechanisms}}
& \multicolumn{3}{c}{\textbf{Metrics}} \\ 
\cmidrule(lr){2-4} \cmidrule(lr){5-7}
& \textbf{Vis.} & \textbf{Chan.} & \textbf{Fid.}
& \textbf{Updates} & \textbf{Latency} & \textbf{Speedup} \\ 
\midrule
Polling
& $\times$ & $\times$ & $\times$
& 1.00$\times$ & 210.0 ms & 1.0$\times$ \\ 

+ Vis. 
& \checkmark & $\times$ & $\times$ 
& 0.54$\times$ & 92.6 ms & 2.3$\times$ \\

+ Chan. 
& \checkmark & \checkmark & $\times$ 
& 0.23$\times$ & 45.7 ms & 4.6$\times$ \\

\addlinespace
\textbf{QUASAR full} 
& \checkmark & \checkmark & \checkmark 
& \textbf{0.11$\times$} & \textbf{20.5 ms} & \textbf{10.2$\times$} \\
\bottomrule
\end{tabular}
\end{table}

In Table~\ref{tab:runtime_ablation}, Vis. denotes visibility pruning, Chan. denotes channel event filtering, and Fid. denotes fidelity event filtering. Updates are normalized to the continuous polling baseline, while speedup is computed from the average update latency. The full QUASAR configuration reduces the normalized update frequency to $0.11\times$ of the polling baseline and achieves a $10.2\times$ latency speedup, showing that the runtime gain comes from the combined event-driven design rather than a single implementation shortcut.

\section{Conclusion}
\label{sec:conclusion}
In this paper, we presented QUASAR, a lightweight spatiotemporal simulator for evaluating entanglement distribution in satellite quantum networks. By abstracting dynamic orbital topologies, time-varying optical transmittance, and quantum memory decoherence into network-layer attributes, QUASAR bridges continuous physical constraints and discrete-event routing evaluation.

Using QUASAR, we modeled two representative satellite paradigms: Simultaneous Downlink and On-Orbit Stitching. We also deployed an EDR-aware routing heuristic as a reference workload to exercise the routing interface. The case study evaluates architecture-dependent EDR, memory-coherence sensitivity, event composition, runtime scalability, trace-driven orbital inputs, concurrent OD workloads, and runtime mechanism ablation. Compared with continuous polling, QUASAR reduces network-layer update latency by over $85\%$, while the ablation study shows that this gain comes from the combined effect of visibility pruning, channel event filtering, and fidelity event filtering. These results suggest that QUASAR provides a practical and extensible framework for future satellite quantum-network protocol evaluation.

\bibliographystyle{splncs04}
\bibliography{mybibliography}
\end{document}